# Tangential deflection and formation of counterstreaming flows at the impact of a plasma jet on a tangential discontinuity

G. Voitcu[1] and M. Echim[1,2]

[1]Institute of Space Science, Magurele, Romania

[2]Belgian Institute for Space Aeronomy, Brussels, Belgium

Corresponding author: Gabriel Voitcu (gabi@spacescience.ro)

**Key Points:**

- The interaction of a magnetosheath plasma jet with a tangential discontinuity with no shear is investigated with 3D-PIC simulations.
- The plasmoid streaming towards the discontinuity is split in two counterstreaming jets drifting tangentially to the magnetopause.
- The deflection and splitting of the jet is produced by a polarization electric field sustained by the charge dependent grad-B drift.

**Abstract**

In this letter we report three-dimensional particle-in-cell simulations of the interaction between a non-penetrable magnetosheath jet and the magnetopause, for northward interplanetary magnetic field. The magnetopause is modeled as a tangential discontinuity with no magnetic shear. We investigate the deflection of the plasma jet in the direction tangential to the magnetopause. We find that as the frontal edge of the jet interacts with the magnetopause, the electrons and ions are scattered in opposite directions, tangential to the magnetopause, by the energy-dependent gradient-B drift. This effect is more effective on the non-thermal particles that tend to accumulate at the two sides of the jet and sustain a polarization electric field in the direction normal to the discontinuity surface. The electric drift of the bulk of particles under the action of this polarization electric field explains the deflection and counterstreaming at the impact of the plasma jet on the tangential discontinuity.

**1 Introduction**

The Earth's magnetosheath is frequently populated by localized plasma structures (called also plasma irregularities, elements, clouds, blobs, jets, streams, plasmoids, density or dynamic pressure enhancements) that are characterized by an excess of density and/or velocity with respect to the background environment (e.g. *Hietala et al.*, 2012; *Savin et al.*, 2012; *Archer and Horbury*, 2013; *Plaschke et al.*, 2013). Most of these structures are propagating anti-sunward and are likely to interact with the magnetopause (e.g. *Dmitriev and Suvorova*, 2012, 2015; *Karlsson et al.*, 2012; *Gunell et al.*, 2014). In-situ data revealed the presence of such magnetosheath irregularities deep inside the magnetosphere (e.g. *Lundin and Aparicio*, 1982; *Woch and Lundin*, 1991, 1992; *Yamauchi et al.*, 1993; *Lu et al., 2004*; *Gunell et al.*, 2012; *Shi et al.*, 2013; *Lyatsky et al.*, 2016). The mechanisms proposed to explain the transfer of mass, momentum and energy at the Earth's magnetopause are not completely elucidated and supplemental investigations are required, particularly for the northward orientation of the interplanetary magnetic field (IMF) (see *Wing et al.*, 2014 and references therein).

The recent study of *Plaschke et al.* (2016) emphasized a significantly large number of magnetosheath high-speed jets impacting the frontside region of the Earth's magnetopause. The impact rate derived from Time History of Events and Macroscale Interactions during Substorms (THEMIS) observations varies from 3 to 9 jets per hour. These high-speed plasma jets can be geoeffective, producing various effects on the magnetospheric environment (*Plaschke et al.*, 2009; *Hietala et al.,* 2012). Their geoeffectiveness depends on the amount of mass, momentum and energy transported towards the magnetosphere. Thus, understanding the dynamics of such plasma structures during the interaction with the magnetopause is very important.

The interaction of localized plasma clouds/jets with transverse magnetic fields has been investigated in the past by several numerical experiments (see, for instance, *Echim and Lemaire*, 2000 and references therein). In a magnetospheric context, the transport and entry of plasma irregularities across tangential discontinuities (TDs), as the Earth's magnetopause, has been simulated with two-dimensional magnetohydrodynamic (MHD) and hybrid codes (*Ma et al.*, 1991; *Dai and Woodward*, 1994; *Savoini et al.*, 1994; *Huba*, 1996). All these studies have shown that 2D magnetosheath plasma filaments with an excess of momentum can propagate without difficulty across the magnetopause and inside the magnetosphere for parallel configurations of the magnetic field, as described theoretically by *Schindler* (1979). It has been also emphasized that the penetration process is independent on the gradient of the magnetic field and that even plasma filaments with small convection velocities can entry into the magnetosphere.

We recently performed three-dimensional particle-in-cell (PIC) simulations to study the interaction of plasma clouds/jets with parallel non-uniform magnetic fields (*Voitcu and Echim*, 2016). Our simulations revealed that the transport and entry of 3D plasma irregularities across TDs is mainly controlled by the dynamic and kinetic pressure of the incoming cloud, its electric polarizability and the height of the magnetic barrier. It has been shown that for a given jump of the magnetic field at the magnetopause, the plasmoids with large enough dynamical pressure cross the TD and move into the magnetosphere, as predicted by the impulsive penetration model (*Lemaire*, 1977). On the other hand, plasma jets with small inertia are completely stopped at the magnetopause and do not entry into the magnetosphere.

The present letter focuses on supplemental kinetic effects observed during the interaction of non-penetrable plasma jets with strong magnetic barriers. We consider an idealized, yet relevant, magnetic configuration corresponding to a magnetosheath high-speed jet impacting the frontside magnetopause during northward IMF. We analyze here the tangential deflection of the plasma jet along the direction normal to both the initial convection velocity and the background magnetic field. The following two science questions are addressed by our study: (i) *What is the physics behind the splitting of non-penetrable plasma jets and the formation of counterstreaming flows tangential to the magnetopause?* and (ii) *What are the typical signatures of such plasma structures that might be observed in-situ by multispacecraft missions?*

## 2 Setup

The numerical simulations are performed with a modified version of the TRISTAN PIC code (*Buneman*, 1993) adapted to study the interaction of localized plasma structures with non-uniform transverse magnetic fields. The code is three-dimensional to allow the simultaneous investigation of the plasma electrodynamics in all relevant directions. A detailed description of the PIC code is given by *Voitcu* (2014).

The simulation setup is illustrated schematically in Figure 1 and is similar to the one used by *Voitcu and Echim* (2016). Figure 1(a) shows the profile of the background magnetic field, $B_0(x)$. The horizontal dashed line marks the critical magnetic field value for which the forward motion along the injection direction is stopped (*Lemaire*, 1985). The panels (b) and (c) of Figure 1 illustrate the initial position and density of the electrons/ions in the planes perpendicular and parallel to the background magnetic field.

The background magnetic field is parallel to the positive *z*-axis and its intensity increases linearly from $B_1$ (in the upstream region, for $x<x_1$) to $B_2$ (in the downstream region, for $x>x_2$). This magnetic field profile corresponds to a thin, steep and impenetrable tangential discontinuity, discussed in *Voitcu and Echim* (2016). The plasma cloud is characterized by a high dielectric constant ($\varepsilon=500$) and its beta parameter is small ($\beta=0.1$, including both dynamic and thermal plasma pressure). The jet is injected in vacuum with an initial bulk velocity of the electrons and ions equal to $U_0=1.2V_{Ti}$ (with $V_{Ti}$ the ion thermal speed). The total simulation time covers 4 ion Larmor periods or equivalently 145 electron Larmor periods.

All physical quantities are normalized as follows: the number density and bulk velocity are normalized to their corresponding initial values $n_0$ and $U_0$. The electric and magnetic fields are normalized to $E_0=U_0 \cdot B_1$ and $B_1$. The time and spatial coordinates are normalized to the initial ion Larmor period $T_{Li}$ and grid spacing $\Delta x=0.4r_{Li}$ (with $r_{Li}$ the Larmor radius of the thermal protons in the upstream region).

## 3 Results

The time evolution of the plasma cloud during the initial stage of the interaction with the tangential discontinuity is shown in Figure 2 and illustrates the electron number density (a-c, same for ions) and the tangential component, $U_y$, of the plasma bulk velocity (d-i), at $t=1.0T_{Li}$ (left column), $t=1.5T_{Li}$ (middle column) and $t=1.9T_{Li}$ (right column), for $z=253$ (a-f) and $x=74$ (g-i) cross-sections. The plasma bulk velocity, $U$, is computed from average velocities of the electrons and ions, as given by equation (2) in *Voitcu and Echim* (2016). To avoid any unrealistically large bulk velocities that could arise in those spatial bins populated with less particles, $U$ is calculated only for those grid cells with a density of at least 5% of $n_0$.

The plasma cloud does not cross the TD (the "magnetopause"). Moreover, the cloud is repelled and pushed back along the negative *x*-axis (in the "magnetosheath"), away from the magnetopause. Indeed, at $t=1.0T_{Li}$ the frontal edge of the plasma cloud is located at $x \approx x_2 = 85$, while at $t=1.9T_{Li}$ the frontal edge is located at $x \approx x_1 = 79$ (panels a and c in Figure 2). At the same time, the plasma element expands rapidly along the magnetic field, diminishing significantly its density.

As discussed by *Voitcu and Echim* (2016), the interaction of the plasma element with the discontinuity is controlled by the magnetic field at the downstream region. If $B_2$ is too large, the forward motion is stopped − the magnetic barrier is closed in this case. It has been shown that this braking is due to the conversion of the bulk motion into gyration motion (adiabatic braking, *Demidenko et al.*, 1967; *Lemaire*, 1985). In the simulations discussed here the initial inertia of the plasma jet would allow propagation in an increasing magnetic field with the intensity less than $B_c=1.06B_1$. However, the magnetic field intensity at the right hand side of the discontinuity is $B_2=1.50B_1$ (Figure 1a). Thus, $B_2>B_c$ or, in other words, the magnetic barrier is too high and the jet is braked before full entry into the magnetospheric side of the discontinuity.

The tangential dynamics of the plasma cloud shows additional interesting features. As the plasma element interacts with the TD, it is deflected quasi-symmetrically, tangential to the discontinuity, along both positive and negative directions of the *y*-axis. Indeed, a non-zero $U_y$ component of the plasma bulk velocity is evidenced at the lateral edges of the cloud, namely for small values of *y*, $U_y<0$ and for large values of *y*, $U_y>0$ (see panels d-i in Figure 2). More exactly, at $t=1.5T_{Li}$ and $z=253$, at larger distance from the discontinuity ($x=74$), the tangential velocity, $U_y$, is equal to $-1.21U_0$ in $y=95$ and $1.48U_0$ in $y=165$, respectively. A little bit closer to the discontinuity ($x=78$), the tangential bulk velocity is equal to $0.07U_0$ in $y=128$. Thus, the magnitude of the plasma bulk velocity in the tangential direction increases as we move away from the center of the cloud towards its lateral edges and is independent of *z* (see panels g-i in Figure 2). An important feature is that both electrons and ions are scattered likewise along $\pm Oy$, suggesting that the driver of the deflection is charge independent.

In order to understand the physical mechanism responsible for this deflection effect, we analyzed in more detail the electrodynamics of the plasma element when it interacts with the magnetic barrier. Figure 3 illustrates the longitudinal, $E_x$, component of the electric field (panels a-b) and the tangential, $U_{E,y}$, component of the zero-order drift velocity (panels c-d), at $t=1.5T_{Li}$, for the same cross-sections as those illustrated in Figure 2. Both $E_x$ and $U_{E,y}$ are shown only for those grid cells populated by enough particles (at least 5% of $n_0$). In panel (e) we plot the variation of $U_y$ (red) and $U_{E,y}$ (blue) as a function of *y*, at $t=1.5T_{Li}$ and for $x=74$. The two profiles have been averaged from $z=150$ to $z=350$ to reduce the numerical noise.

One can notice in Figure 3(a-b) the existence of two distinct regions or "wings" at the lateral edges of the plasma cloud that exhibit opposite polarization of the longitudinal component of the electric field. On the one hand, $E_x$ takes preponderantly positive values at the lower edge of the plasma cloud (for $y<110$) and, on the other hand, $E_x$ takes preponderantly negative values at the upper edge of the cloud (for $y>150$). The zero-order drift velocity due to this longitudinal electric field is oriented along $-Oy$ direction for $y<110$ (where also $U_y<0$) and along $+Oy$ direction for $y>150$ (where also $U_y>0$). $U_y$ and $U_{E,y}$ show similar features as results from a comparison of Figure 2(e-h) with Figure 3(c-d) and emphasized in Figure 3(e). These results indicate that the tangential deflection of the plasma cloud is produced by the longitudinal electric field, $E_x$, via the charge independent zero-order drift, $U_{E,y}=-E_x/B_z$.

In Figure 4(a) we plot the net charge density at the boundaries of the plasma cloud in the perpendicular plane to the background magnetic field. The charge density has been averaged along the magnetic field lines from $z=179$ to $z=327$ to diminish the intrinsic PIC noise controlled by the limited number of particles loaded into the simulation domain. In order to better illustrate the polarization of the cloud's edges, we show the mean net charge density inside its core.

The frontal edge of the plasma element ($x\approx 82$) is electrically polarized. Indeed, a negative space charge layer is observed at the lower wing of the propagation front (for $y<110$), while a positive layer is evidenced at the upper lateral wing (for $y>150$). The position of the space charge layers is reversed in the body of the jet that did not interact yet with the magnetic discontinuity; for $x<74$, the lower lateral edge ($y\approx 70$) is positively polarized, while the opposite one ($y\approx 175$) is negatively charged. The space charge layers in the front side of the jet ($x\approx 82$) and the ones in the body of the jet ($x<74$) sustain the longitudinal electric field, $E_x$, discussed above.

The charge layers at the jet's propagation front are built by the continuous action of the charge dependent gradient-B drift in the transition region. As the first parcels of the jet enter the magnetic discontinuity, the electrons and ions are scattered by the gradient-B drift in opposite directions along the *y*-axis with a velocity proportional to the gyration energy. Thus, the most energetic particles are scattered more efficiently than the thermal ones. During one ion Larmor period, the frontside suprathermal ions with a gyration velocity of $1.5V_{Ti}$ drift along $+Oy$ axis over a distance of $3\Delta x$, while the suprathermal electrons drift in the opposite direction over a distance of $75\Delta x$, comparable with the initial width of the plasma jet. Therefore, the charge dependent first-order drift separates ions in the upper wing region (at larger *y*-values) and electrons in the lower wing region (at smaller *y*-values), as illustrated in Figure 4(b). These space charge layers persist and develop in time since the particles do not have enough energy to penetrate the magnetic barrier. Note that the effect of the gradient-B and electric drifts cannot be disentangled. The two act together and contribute to the global deflection of the plasma cloud in the tangential direction and to the formation of counterstreaming plasma flows. It should also be mentioned that in our simulations the polarization drift is significantly smaller than the gradient-B drift.

The later stages of the simulation are shown in the top panels of Figure 5 and illustrate the electron number density (panel a), the tangential component, $U_y$, of the plasma bulk velocity (panel b) and the tangential component, $U_{E,y}$, of the zero-order drift velocity (panel c), at $t=4.0T_{Li}$, for $z=403$. The plasma element is finally disrupted in two parts that continue to counterdrift quasi-symmetrically along $Oy$, while they are also slightly pushed away from the TD. After four ion Larmor periods from the beginning of the simulation, the lateral width of the plasma element covers ~90 ion Larmor radii (~5 times wider than initially). Thus, the magnetosheath plasma cloud streaming towards the magnetic discontinuity has been split into a jetting structure that drifts in opposite directions tangentially to the magnetopause with a velocity of ~$1.5U_0$, larger than the impact velocity, $U_0$.

Such tangentially deflected plasmoids might be detected *in-situ* by multi-spacecraft missions flying in the vicinity of the Earth's magnetopause, as, for instance, Cluster, THEMIS or Magnetospheric Multiscale (MMS) probes. In order to estimate a typical signature from satellite data of such non-penetrable plasmoids/jets, we "launched" four virtual satellites into our simulation domain in the vicinity of the tangential discontinuity. The satellites "measure" the particle density and plasma flow velocity during four ion Larmor periods of runtime.

The "time series" data collected by the virtual satellites are illustrated in the bottom panels of Figure 5 that show the electron number density (panel d) and velocity components in the directions normal ($U_x$, panel f) and tangential ($U_y$, panel e) to the magnetopause, respectively. The four virtual satellites are located at $x$=75, $y_1$=153 (satellite S1), $y_2$=186 (satellite S2), $y_3$=220 (satellite S3), $y_4$=253 (satellite S4) and $z$=403. We consider the satellites fixed, since a realistic satellite velocity (~4 km/s) is much smaller than the plasmoid's velocity and the corresponding satellite displacement during $4T_{Li}$ of runtime is smaller than the spatial resolution of our simulation. The satellite measurements are taken with a temporal sampling step of $0.1T_{Li}$.

The plasmoid is detected shortly after injection by the density measurements of the inner probes (S2 and S3) located close to the right edge. The outer probes (S1 and S4), located farther away, detects the plasma cloud at later times and the density signature is fainter. The particle density is decreasing rapidly due to the fast parallel expansion along the magnetic field, but it varies also with the distance from the cloud's boundaries where $n_e$ drops sharply. Outside the jet, the satellites measure the void, as the particles are injected in vacuum. The gradual decrease of the normal velocity, $U_x$, observed by the satellites S2 and S3 during the first ion Larmor period of sampling indicates the deceleration of the plasma flow during the interaction with the discontinuity. At $t$≈$1.25T_{Li}$, a flow reversal ($U_x$<0) is detected by the inner probes (S2 and S3) in the direction normal to the magnetopause. Simultaneously, the outer probes (S1 and S4) measure a very fast tangential flow ($|U_y|$≈$2.5U_0$) together with a backward motion along the *x*-axis. The four spacecraft measure oppositely directed plasma jets: the spacecraft S1 and S2 detect a negative tangential velocity, while the spacecraft S3 and S4 detect a positive one.

The virtual time series illustrated in Figure 5 have been obtained for a multispacecraft configuration parallel to the discontinuity surface that covers a large region of the plasma structure. Thus, depending on the actual satellites configuration and the interprobe distance, the measured times series can be more or less similar to the ones presented here. Due to the larger satellites separation, it is more likely for Cluster and THEMIS to detect the particular signatures identified by our simulations.

**4 Conclusions**

In this paper we used a 3D-PIC code to simulate the interaction of a plasmoid or jet with the magnetic field of a tangential discontinuity in an idealized configuration corresponding to the dayside magnetopause for a northward IMF. We analyzed the dynamics of the plasma in the direction tangential to the discontinuity. Here are the main findings of our simulations:

1. The interaction of the plasmoid with the tangential discontinuity leads to a splitting of the plasma element into two oppositely directed plasma jets drifting tangential to the magnetic discontinuity. The tangential component of the plasma bulk velocity increases towards the lateral edges of the plasmoid.

2. The tangential deflection of the plasma cloud is produced by the zero-order drift of electrons and ions due to the longitudinal polarization electric field formed at the frontal edge of the jet interacting with the magnetic discontinuity.

3. The frontside polarization of the plasma cloud is due to oppositely scattering of electrons and ion by the charge-dependent gradient-B drift. Two space charge layers of reversed polarities are developed in time at the frontside region of the plasmoid. The electric and gradient-B drifts act simultaneously and contribute together to the global deflection of the plasma cloud in the $\boldsymbol{U}_0 \times \boldsymbol{B}_0$ direction. The longitudinal polarization electric field reported here is a kinetic, non-MHD, effect that demonstrates the important role of Larmor scale processes for plasma dynamics in non-uniform electromagnetic fields.

4. "Data" from four virtual satellites "launched" into the simulation domain illustrate possible fingerprints observable in the *in-situ* measurements of plasma density and flow velocity. We identified the following characteristic features in the "time series" of the quantities "observed" by the virtual satellites: (i) a reversal of the plasma flow in the direction normal to the magnetopause ($U_x$ decreases and then turns negative), (ii) a very fast tangential flow observed simultaneously with the reversal of $U_x$ and (iii) two oppositely directed tangential flows ($U_y$ changes sign and magnitude along *y*-axis) observed by satellites positioned symmetrically with respect to the center of the jet.

The new and original result of our study is the identification of a physical mechanism based on the guiding center approximation that leads to splitting of the non-penetrating plasma jets and the formation of counterstreaming flows tangential to the magnetopause surface. Such non-penetrating magnetosheath jets have been observed in the vicinity of the magnetopause. The study of *Dmitriev and Suvorova* (2015) shows that approximately 40% of the jets detected by THEMIS in the magnetosheath do not penetrate the magnetopause, while the rest of 60% are able to cross over and enter inside the magnetosphere. The non-penetrating jets have velocities smaller than a given threshold, while the penetrating ones are characterized by larger velocities, thus consistent with our simulations (see also *Voitcu and Echim*, 2016) and with the impulsive penetration mechanism (*Lemaire*, 1977, 1985).

The mechanism we propose can explain the development of jetting plasma structures in the vicinity of the frontside magnetopause during northward IMF. The typical satellite signature of such plasma structures obtained from our simulations can be further used to identify the non-penetrable jets deflected at the Earth's magnetopause by using *in-situ* data from multispacecraft missions as Cluster, THEMIS or MMS.

**Acknowledgments and Data**

The authors acknowledge support from the European Community's Seventh Framework Programme through grant agreement 313038/2012 (STORM). The simulation data used to produce all the plots included in this paper can be requested by sending an e-mail to Gabriel Voitcu at one of the following addresses: gabi@spacescience.ro or gabriel.voitcu@gmail.com.

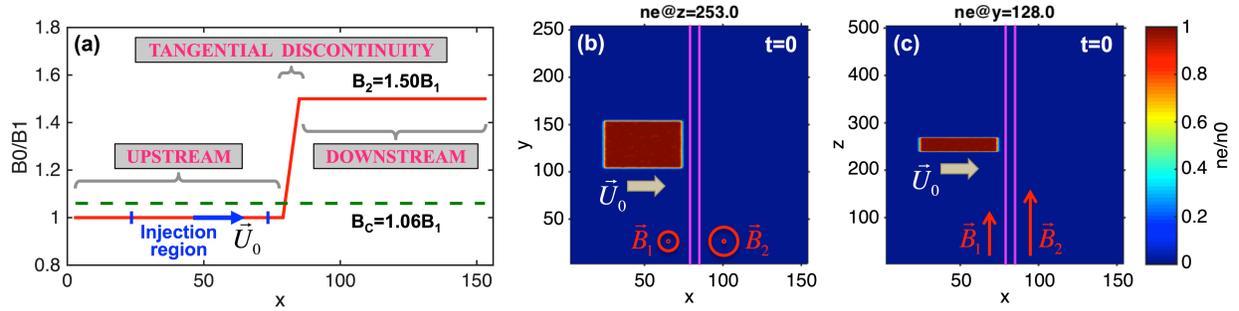

**Figure 1.** (a) The profile of the background magnetic field; the background field is parallel to +$Oz$ and its intensity increases by 50% over 2.3 ion Larmor radii. The initial density of electrons in the planes perpendicular and parallel to the magnetic field for $z$=253 and $y$=128 are shown in panels (b) and (c). The two magenta lines mark the boundaries of the TD.

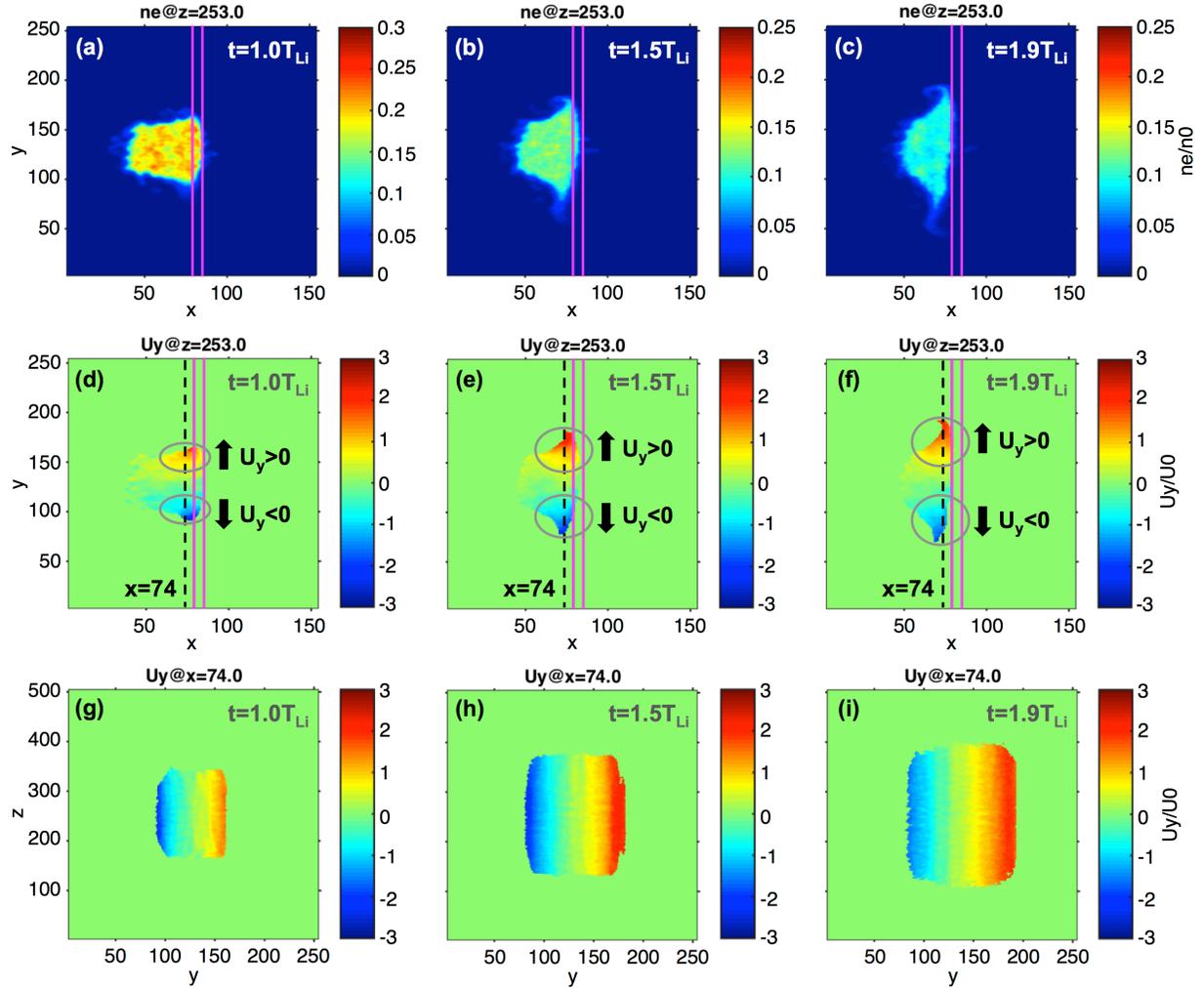

**Figure 2.** Panels (a)-(c) show a 2D section in *xOy* plane (at *z*=253) of the electron number density (same for ions) for three different simulation instants: $t=1T_{Li}$, $t=1.5T_{Li}$ and $t=1.9T_{Li}$. The panels (d)-(f) show a 2D section of the tangential component of the plasma bulk velocity, $U_y$, in the plane *xOy*, for *z*=253 and the same three moments of time as above. The panels (g)-(i) show $U_y$ in the *yOz* plane at *x*=74.

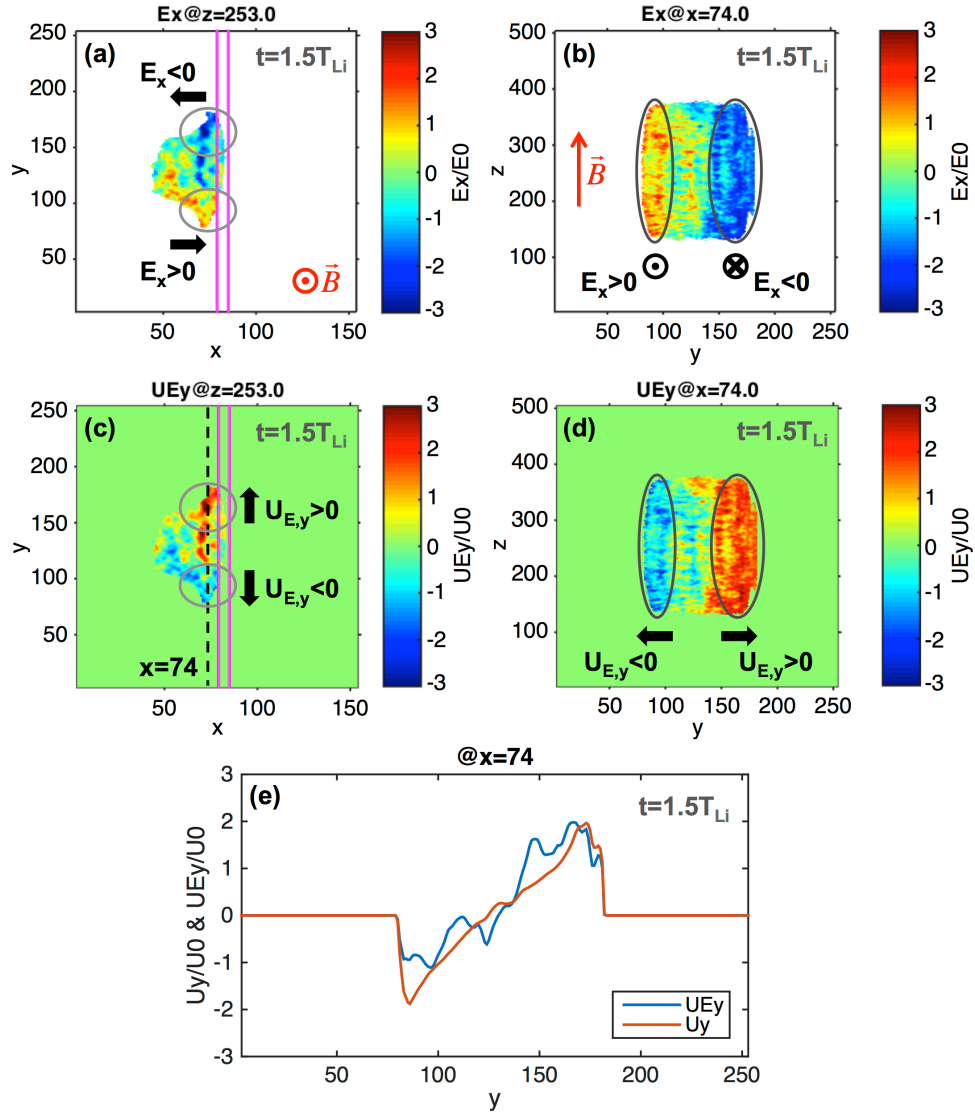

**Figure 3.** (a)-(b) $E_x$ component of the electric field and (c)-(d) $U_{E,y}$ component of the electric drift velocity at $t=1.5T_{Li}$, for (left column) $z=253$ and (right column) $x=74$ cross-sections inside the simulation domain. (e) Plasma bulk velocity (red) and electric drift velocity (blue) variation profiles along the $Oy$ direction at $t=1.5T_{Li}$ and for $x=74$.

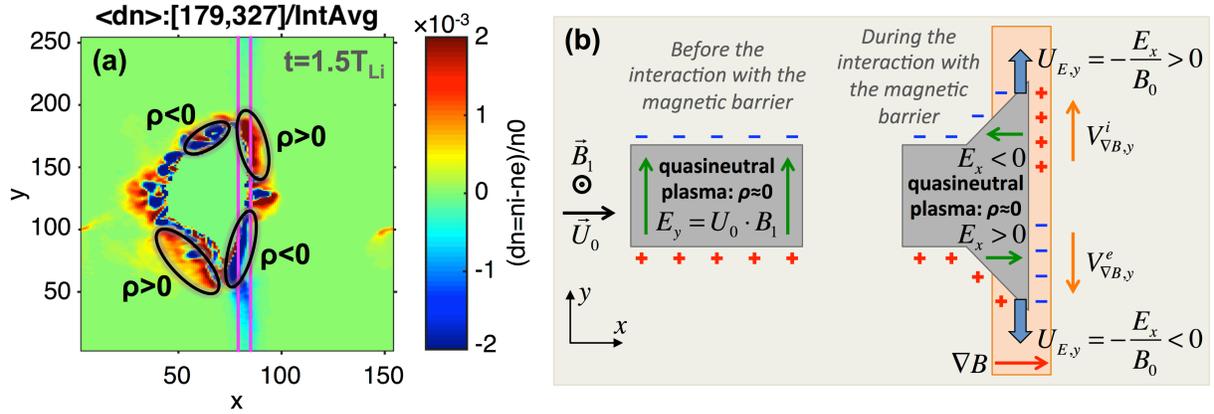

**Figure 4.** (a) Total charge density in the perpendicular plane to the magnetic field, at $t=1.5T_{Li}$, averaged from $z=179$ to $z=327$. (b) Schematic diagram illustrating the longitudinal polarization of the plasma element while interacting with the non-uniform transverse magnetic field.

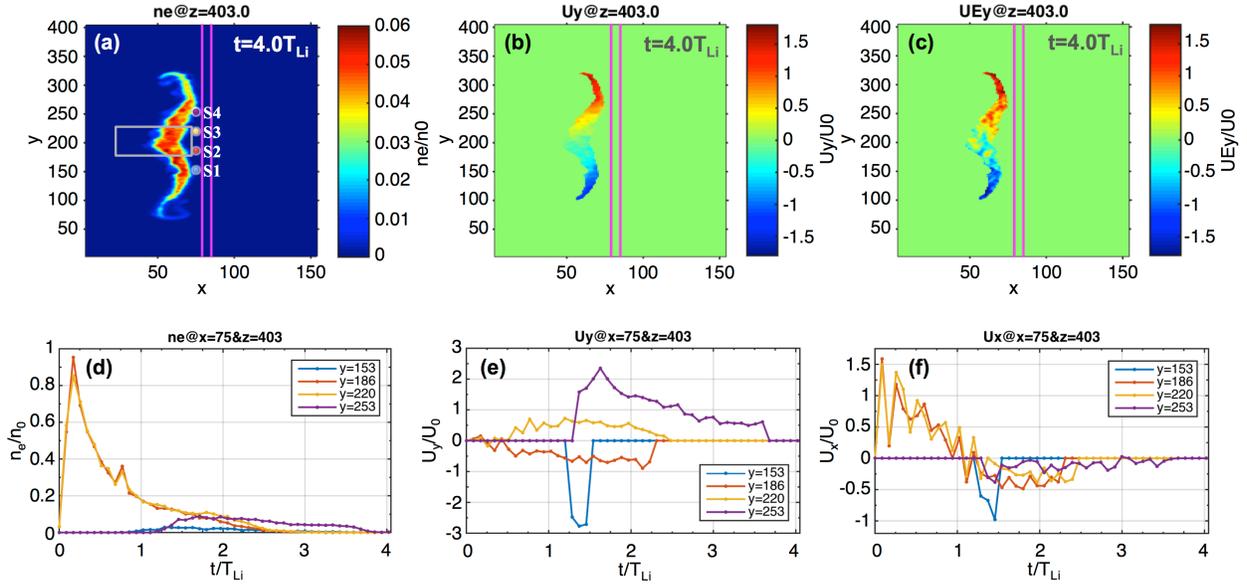

**Figure 5.** (a) Electron number density (same for ions), (b) $U_y$ component of the plasma bulk velocity and (c) $U_{E,y}$ component of the zero-order drift velocity, at $t=4.0 T_{Li}$, for $z=403$. "Time series" from the four virtual probes (S1, S2, S3, S4) in panel (a) show: (d) electron number density, (e) $U_y$ and (f) $U_x$ components of the plasma bulk velocity as "measured" by the virtual satellites located at $x=75$, $y_1=153$ (S1), $y_2=186$ (S2), $y_3=220$ (S3), $y_4=253$ (S4) and $z=403$. The grey rectangle in panel (a) marks the boundaries of the plasmoid at $t=0$.